\documentclass[aps,12pt,preprintnumbers,nofootinbib,superscriptaddress]{revtex4}
\usepackage[letterpaper]{geometry}                
\usepackage{graphicx}
\usepackage{amssymb,amsmath}
\usepackage{epstopdf}
\begin{document}

\newcount\hour \newcount\minute
\hour=\time \divide \hour by 60
\minute=\time
\count99=\hour \multiply \count99 by -60 \advance \minute by \count99
\newcommand{\mydate}{\ \today \ - \number\hour :00}

\preprint{CALT 68-2660}

\title{The Higgs Decay Width in Multi-Scalar Doublet Models }

\author{Sonny Mantry}
\email[]{mantry@theory.caltech.edu}
\affiliation{California Institute of Technology, Pasadena, CA 91125}

\author{Michael Trott}
\email[]{mrtrott@physics.ucsd.edu}
\affiliation{Department of Physics, University of California at San Diego, La Jolla, CA 92093}

\author{Mark B. Wise}
\email[]{wise@theory.caltech.edu}
\affiliation{California Institute of Technology, Pasadena, CA 91125}


\begin{abstract}
We show that there are regions of parameter space in multi-scalar doublet models where, in the first few hundred inverse femtobarns of data, the new charged and neutral scalars are not directly observable at the LHC and yet the Higgs decay rate to $b \,  \bar b$ is changed significantly from its standard model value. For a light Higgs with a mass less than $140 ~{\rm GeV}$, this can cause a large change in the number of two photon and $\tau^+ \tau^-$ Higgs decay events expected at the LHC compared to the minimal standard model. In the models we consider, the principle of minimal flavor violation is used to suppress flavor changing neutral currents. This paper emphasizes the importance of measuring the properties of the Higgs boson at the LHC; for a range of parameters the model considered has new physics at the TeV scale that is invisible, in the first few hundred inverse femtobarns of integrated luminosity at the LHC, except indirectly through the measurement of Higgs boson properties.
\end{abstract}

\maketitle
\newpage
\section{Introduction}

 Experiments at the LHC  will directly probe physics at the weak scale. Most physicists believe that there is new physics at this energy scale, beyond what is in the minimal standard model (SM). This belief is motivated to a large extent by the hierarchy puzzle and by the fact that the scalar sector of the standard model has yet to be directly probed by experiment. The single doublet in the SM is  the simplest example of a scalar sector but many extensions have been studied. Amongst the most widely considered are two Higgs doublet models \footnote{See \cite{Reina:2005ae} for a review.}. A problem that immediately arises in such models is the possibility of flavor changing neutral current (FCNC) effects that are unacceptably large. In particular, when the SM fermion fields couple to both doublets,  and the couplings are 
arbitrary, FCNC effects are possible at tree level. Glashow and Weinberg gave a simple prescription for how to avoid such effects through imposing a discrete symmetry \cite{Glashow:1976nt}.  One can also suppress FCNC effects by adopting an ansatz suppressing the coupling of the new doublet \cite{Cheng:1987rs,Luke:1993cy,Antaramian:1992ya,Atwood:1996vj}.  In this paper, we use the principle of minimal flavor violation  (MFV) \cite{Chivukula:1987py,Hall:1990ac,D'Ambrosio:2002ex,Cirigliano:2005ck,Buras:2003jf,Branco:2006hz} which causes 
tree level FCNC to vanish (or at least be suppressed by small mixing angles) in multi doublet models in a natural way.

We are interested in the possible effects of new physics on the properties of the Higgs boson and we characterize the impact of new physics on the Higgs through an operator analysis \cite{Manohar:2006gz,Grinstein:2007iv,Kile:2007ts,Graesser:2007yj,Mantry:2007sj,Pierce:2006dh,Fox:2007in} . We assume that the new physics mass scale $M$ is much larger than the Higgs boson mass and add higher dimension operators that are invariant under the $SU(3)\times SU(2)\times U(1)$ symmetry of the standard model. Since the operators give small corrections to the SM, one does not expect them to influence standard model processes that are unsuppressed. For example, the Higgs coupling to two W bosons is an unsuppressed tree level coupling and new physics contributions to it should be negligible. However the dominant Higgs production mechanism through gluon fusion, $g \, g \rightarrow h$ occurs at leading order in perturbation theory through a top quark loop. Hence, in the SM it is suppressed and new physics can easily compete with the standard model contribution~\cite{Manohar:2006gz}. Similar remarks hold for the $h \rightarrow \gamma \gamma$ decay amplitude. Some of the tree level couplings of the standard model Higgs are also very small. For example, the Higgs to $\tau$ Yukawa coupling is of order\footnote{Here $v \simeq {\rm 250~GeV}$ is the vacuum expectation value that spontaneously breaks the weak gauge group down to the electromagnetic gauge group} $m_{\tau}/v \sim 0.75 \times10^{-2}$ and the Higgs to b-quark Yukawa coupling is of order $m_b/v \sim 2\times 10^{-2}$. New physics characterized by higher dimension operators can also compete with the standard model in the $h\rightarrow \tau \, \bar \tau$ ~\cite{Mantry:2007sj} and $h \rightarrow b \, \bar b$ decay amplitudes.

A light Higgs with a mass less than $140~{\rm GeV}$ is likely to be detected first through its decay to two photons despite the fact that the branching ratio for this process is quite small, {\it i.e.} of order $10^{-3}$. Early detection through its decay to $\tau^+ \tau^-$, which has a branching ratio around $10^{-1}$, may also be possible. The dominant decay mode is to $b \, \bar{b}$ pairs. However this decay mode is much harder to observe because of large theoretical uncertainties\footnote{See \cite{Rainwater:2007cp} for a recent review on production and detection of the Higgs and \cite{Benedetti:2007sn} for a recent study on the $t \, \bar{t} \, b \, \bar{b}$ SM background.} on the cross section of the irreducible SM background $t \, \bar{t} \, b \, \bar{b}$. An integrated luminosity of order $500 \, {\rm fb}^{-1}$ may be required to observe the standard model Higgs in the $b \, \bar b$ channel. If the rate for $h \rightarrow b  \, \bar b$ decay is changed by a factor $f$ from its standard model value (but other properties of the Higgs are left unaltered) then the number of $h \rightarrow \gamma \gamma$ or $h \rightarrow \tau^+ \tau^-$ decay events observed at the LHC is changed by a factor $\xi$ where
\begin{eqnarray}
\xi = \frac{1}{1+ (f-1)\text{Br}(h\to b\bar{b})_{\rm SM}},
\end{eqnarray}
and $\text{Br}(h\to b\bar{b})_{\rm SM}$ is the SM branching fraction of $h\to b\bar{b}$.
 Even though $h \rightarrow b \, \bar b$ decay will not be directly observable until there are many years of LHC data, it's rate is of crucial importance for all measurable properties of the low mass Higgs.

In this paper, we consider multi-doublet scalar models.\footnote{New physics in the form of a second scalar doublet with an unbroken $Z_2$ symmetry can also provide a component of dark matter and the effect of such a second doublet on the SM
Higgs was recently examined in \cite{Cao:2007rm}.} For simplicity we restrict our attention to two scalar doublets $H$ and $S$. Here $H$ denotes the usual SM Higgs doublet with a mass less than $140$ GeV.  $S$ is a new scalar doublet with a mass $ M \sim 1$ TeV and has the same quantum numbers as the SM Higgs doublet $H$. We demonstrate that there are regions in parameter space for which the new doublet $S$ will be invisible at the LHC, at least in the first few hundred inverse femtobarns of data. The largest observable effect of this new doublet $S$ will be order one shifts in the $h\to b \, \bar{b}$ rate which in turn will affect the branching ratios for all light Higgs decay channels.  It is straightforward to generalize our results to a scenario with more than two scalar doublets. 

\section{The Two Doublet Model}

The scalar potential for the model we consider is given by  \footnote{We thank Lisa Randall for pointing out the typo in the sign of $g_1$ in Eqn.(2) in the previous version \cite{Randall:2007as}.}
\begin{eqnarray}
&&V(H,S)= {\lambda \over 4}\left(H^{\dagger} H- {v^2 \over 2}\right)^2+M^2S^{\dagger}S +{\lambda_S \over 4} \left(S^{\dagger}S \right)^2-\left[g_1\left( S^{\dagger}H \right)\left( H^{\dagger}H \right) +{\rm h.c.} \right] \nonumber \\
&&+g_2\left( S^{\dagger}S \right)\left( H^{\dagger}H \right)+\left[g_2'\left( S^{\dagger}H \right)\left( S^{\dagger}H \right)+{\rm h.c.} \right]+g_2''\left(S^{\dagger} H \right)\left(H^{\dagger}S\right) \nonumber \\
&&+\left[g_3\left( S^{\dagger}S \right)\left( S^{\dagger}H \right)+{\rm h.c.} \right].
\end{eqnarray}
We assume the phase of $S$ is adjusted so that $g_1$ is real.
$S$ appears with a positive mass term and acquires a vacuum expectation value only through it's coupling to $H$, which undergoes the usual electroweak symmetry breaking.
Since $M$ is much greater than the weak scale $v$, the neutral component $S^0$ gets a vacuum expectation value that is much smaller than $v$
\begin{equation}
\label{Svev}
\langle S^0 \rangle \simeq {g_1 v^3 \over 2 {\sqrt 2} M^2} \ll v.
\end{equation}

In addition to the SM Yukawa couplings of the doublet $H$, the doublet $S$ has the following Yukawa couplings to the quarks,
\begin{equation}
\label{yuk}
\Delta{\cal L}_Y= -\eta_D{\bar d_R} \tilde{g}_D S^{\dagger}Q_L -\eta_U{\bar u_R} \tilde{g}_U S {\epsilon} Q_L +{\rm h.c.}
\end{equation}
We make use of  the principal of minimal flavor violation. This results in small loop level FCNC through the appearance of Yukawa coupling matrices $\tilde{g}_D$ and $\tilde{g}_U$ in the loops. We assume that possible multiple insertions of the Yukawa matrices in Eq.~(\ref{yuk}) are suppressed.  In this model, the physical quark masses are the result of the sum of contributions from the coupling of quarks to $H$ and $S$. Thus in the mass eigenstate basis, the Yukawa matrices $\tilde{g}_{U,D}$ do not satisfy the usual relation to physical quark masses. In the SM, $g^i_{U,D} = \sqrt{2} \, m_i/v$. In the down quark sector, in the mass eigenstate basis, the couplings of the heavy scalar doublet  are
\begin{equation}
\Delta{\cal L}_Y= -\eta_D {\sqrt 2} {\bar d_R}{\tilde{m}_d \over v} V^{\dagger}u_L S^- -\eta_D {\sqrt 2}  {\bar d_R}{\tilde{m}_d \over v} d_L S^0 +{\rm h.c.}
\end{equation}
Here $V$ is the CKM matrix, $\tilde{g}_D =\sqrt{2}\tilde{m}_d/v$ and $\tilde{m}_d$ is related
to the physical down quark mass $m_d$ by 
\begin{eqnarray}
\label{physicalmass}
\tilde{m}_d=\frac{m_d}{(1+\sqrt{2}\eta_D\langle S^0 \rangle/v)} .
\end{eqnarray}
 
We assume that the constant $\eta_U$ in Eq. (\ref{yuk}) is very small so that the $S$ coupling to the up-type quarks can be neglected.  When $\eta_U \ll 1$ the production of $S$ via it's coupling to the top quarks is suppressed. 

On the other hand, we take $\eta_D$ to be large, $\eta_D \sim 10$ and for simplicity we choose it to be real. Since the down type quark Yukawa couplings in the matrix $\tilde{g}_D$ are very small, the effective coupling $\eta_D  \,  \tilde{g}_D$ is still perturbative.  The choice of $\eta_D \gg 1$ makes the coupling of  $b$ quarks stronger to
$S$ compared to $H$ resulting in a large shift in $h\to b \, \bar{b}$ once $S$ is integrated out. 
Thus, with $\eta_U \ll 1$, $\eta_D \gg 1$, and $M\gg v$ an almost invisible $S$ can be produced
at the LHC and will leave it's footprint through large shifts in the $h\to b \, \bar{b}$ rate.

\section{FCNC Constraints}

\begin{figure} 
\centerline{\scalebox{1.0}{\includegraphics{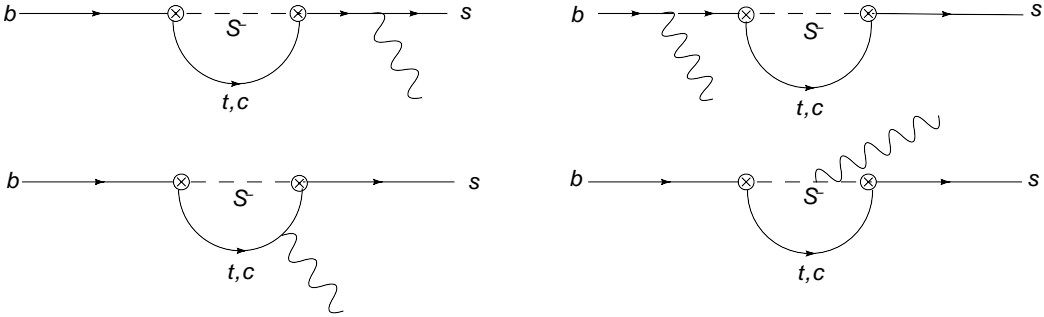}}}
\caption{The one loop contribution to $b \rightarrow s \, \gamma$ due to the doublet S.}
\label{bsgamma}
\end{figure}

Even though we imposed MFV eliminating the possibility of tree level FCNC, in our two doublet model, there are at one loop corrections to standard model FCNC processes. In particular, we want to check that the choice of $|\eta_D| \sim 10 \gg 1 $ is consistent with constraints from FCNC. Consider the weak radiative b decay $b \rightarrow s \gamma$ with two doublets \cite{Grinstein:1987pu}. Calculating the Feynman diagrams in Fig.(\ref{bsgamma}) we find that charged $S$ exchange induces at one loop the effective Hamiltonian,
\begin{equation}
\label{hamiltonian}
{\cal H}_{\rm eff}= {e \over 96 \pi^2}\eta_D^2 {m_t^2 \over M^2}{G_F \over {\sqrt 2}} V_{ts}^*V_{tb} \Big(\frac{\tilde{m}_b}{m_b}\Big )\tilde{m}_s{\bar s}_R \sigma_{\mu \nu} F^{\mu \nu} b_L,
\end{equation}
where $e< 0$ is the electron charge.

For inclusive decay, this does not interfere with the standard model contribution from $O_7$ \cite{Grinstein:1987vj} since the strange quarks have opposite chirality. Hence, we find
\begin{equation}
\label{radiativerate}
{ \Gamma(b \rightarrow s \gamma) \over \Gamma(b \rightarrow s \gamma)_{\rm SM}}\simeq 1+\left({\eta_D^2 \, \tilde{m}_s \, \tilde{m}_b \, m_t^2 \over 24 \, C_7(m_b) \, m_b^2 \, M^2}\right)^2.
\end{equation}
Taking $\eta_D=10$, $g_1=1$, $M=1~{\rm TeV}$ and  $| C_7(m_b) | \simeq 0.3$ in Eq.~(\ref{radiativerate}) gives $ \Delta \Gamma(b \rightarrow s \gamma) / \Gamma(b \rightarrow s \gamma)_{\rm SM} \sim 10^{-5}$, which is much too small to be observed\footnote{See \cite{Misiak:2006zs,Misiak:2006ab} for the latest calculation of BR($\bar{B} \to X_s \, \gamma$) at NNLO in QCD and its comparison to the results from Babar \cite{Aubert:2005cua,Aubert:2006gg} and Belle \cite{Koppenburg:2004fz} as averaged by the Heavy Flavor Averaging Group \cite{Barberio:2006bi}. The remaining uncertainties in theory and experiment preclude a exclusion of our model based on $b \rightarrow s \, \gamma$ constraints for the parameter space of interest.}. For exclusive decays there can be interference between the standard model contribution and the contribution of the effective Hamiltonian in Eq.~(\ref{hamiltonian}) but there hadronic uncertainties cloud our ability to constrain the new physics \cite{Becirevic:2006nm,Ball:2006eu}.

There are contributions from the interactions of the new doublet $S$ to the Wilson coefficient $C_7$ that are proportional to $\eta_D \eta_U$ and are not suppressed by ${\tilde m}_s /m_b$. However, because we have focused on the region of parameter space where $\eta_U$ is very small these have been omitted. Our assumption of very small $\eta_U$ greatly diminishes the constraint that $b \rightarrow s \gamma$ places on the model.

\section{Effects on light Higgs decays}\label{effects}
So far we have described the general features of the two doublet model we are considering and demonstrated the  compatibility of a large $|\eta_D| \sim 10$ with $b \rightarrow s \, \gamma$ constraints. Next we investigate how this simple extension of the SM affects light Higgs decay to quarks by integrating the heavy doublet $S$ out of the theory to induce the effective operator
\begin{equation}
\label{effective}
{\cal L}_{\rm eff}=-\eta_D(H^{\dagger}H) {g_1 \over M^2}{\bar d_R} \tilde{g}_D H^{\dagger}Q_L + {\rm h.c.}
\end{equation}
Including the effects of this operator and using Eqs.(\ref{physicalmass})  and (\ref{Svev}) we find that the $h\to b \, \bar{b}$ rate is modified relative to the SM as
\begin{equation}
\label{dramatic}
{\Gamma(h \rightarrow  b \bar b )\over \Gamma(h \rightarrow \bar b b)_{\rm SM}}=\left[ 1+{3 v^2 g_1 \eta_D / 2 M^2} \over 1+ {v^2 g_1 \eta_D / 2 M^2} \right]^2.
\end{equation}
Note we have included terms suppressed by more powers of $v^2/M^2$ in Eq. (\ref{dramatic}) which are accompanied by the large factor $\eta_D$. However, we can still consistently ignore the effects of dimension 8  operators contributing to $h\to b \, \bar{b}$ since their contributions start at order $\eta_D  \, v^4/M^4\ll 1$.

At $M=1$ TeV,  the parameter choices of $g_1=0.5, \eta_D=10$ or  $g_1=1, \eta_D=5$ give 
the rate for $h\to b\bar{b}$ which is 1.6 times it's SM value. An even more dramatic effect is seen for the parameter choices of
$g_1=-2, \eta_D=20$ and $g_1=1, \eta_D=-10$ which give a rate that is 121 and 0.008 times the SM value respectively. 
Thus, the presence of an additional  ${\rm TeV}$ scale scalar doublet with a coupling to $b$ quarks about ten times the SM value can have dramatic changes in the decay width and branching fractions of a light Higgs. For example, with  $g_1=-2, \eta_D=20$ the branching ratio for the experimentally promising modes of $h\to \gamma \gamma$ and $h\to \tau^+ \tau^-$ will be down by a factor of  $\xi \sim 1/80$. Such a scenario would make detection of the light Higgs very difficult. On the other hand for $g_1=1, \eta_D=-10$ in which case the $h\to b\bar{b}$ rate is 0.008 it's SM value. In this case, the branching fractions for $h\to \gamma \gamma$  and $h\to \tau^+ \tau^-$ will increase by the factor $\xi \sim 3$ for $m_h =120$ GeV. This would apply for Higgs searches at the Tevatron as well \cite{Acosta:2005bk}. \footnote{New physics in the form of a massive fourth generation neutrino \cite{Belotsky:2002ym} or additional scalar singlets \cite{BahatTreidel:2006kx} have also been shown to 
effect the possibility of the detection of the Higgs at LHC.}

One can also generalize the above analysis to the lepton sector
and induce a corresponding effective operator
\begin{equation}
\label{effectivelepton}
{\cal L}_{\rm eff}=-\eta_{\ell}(H^{\dagger}H) {g_1 \over M^2}{\bar e_R} \tilde{g}_{\ell} H^{\dagger}L_L + {\rm h.c.} ,
\end{equation}
which contributes to the decay of the Higgs to charged leptons
$h\to \ell^+ \ell^-$. By choosing a large value for $\eta_{\ell}$ one can similarly induce order one shifts in the decay rate to $h\to \ell^+ \ell^-$.
Such order one shifts can be seen at the LHC in the experimentally promising channel of $h\to \tau^+ \tau^-$. The effect of the operator in Eq.(\ref{effectivelepton}) was recently studied in \cite{Mantry:2007sj} 
where naturalness criteria were 
used to constrain the size of the Wilson coefficient. It was shown that order one shifts are indeed possible and compatible with experimental constraints. The branching ratio for $h\to \tau^+ \tau^-$ can be influenced both by the effect of the operator in Eq.~(\ref{effectivelepton}) on the rate for $h\to \tau^+ \tau^-$ and by the effect of the operator in Eq.~(\ref{effective}) on the total Higgs width.


Order one shifts in the rate for $h\to \tau^+ \tau^-$ can also affect the total width of the light Higgs   since it's branching ratio is not negligible.  For example, if $m_h=120$ GeV then the branching ratio in the SM for $h\to \tau^+ \tau^-$ is about $7\%$. As order one corrections are possible to partial decay widths for both $h\to b \bar{b}$ and $h\to \tau^+\tau^-$, the relative impact of the 
two decays on the total width can be changed dramatically.
This can make the total decay width even more sensitive to the effects of $S$.  However, for the sake of simplicity we will assume in this paper that $\eta_{\ell}$ 
is small so that effects on the width of the Higgs from the coupling of $S$ to leptons are negligible.

The number of observed $h \rightarrow \gamma \, \gamma$ events can also be effected by higher dimension operators that 
induce a direct coupling between $h$ and $\gamma \gamma$ and between $h$ and $g g$ as discussed in \cite{Manohar:2006gz}. The latter effects the Higgs production rate by gluon fusion.
New physics effects of this form are distinguishable from a change in the total width as the new physics effects on the 
total width will cancel in the ratios of the number of expected events for different Higgs production mechanisms and decay channels. 


\section{Production and Decay of the new scalar doublet}

We now study the production and decay of the new scalar doublet $S$ and discuss the possibility for it's observation at the LHC. The doublet contains new neutral and charged scalars with masses approximately equal to $M$. If $M  \simeq 1~{\rm TeV}$ the LHC has enough energy to produce these states. However, we will show that for a range of parameters their production rates are quite small and that the dominant decay channels have poor experimental signatures making them invisible at the LHC, at least for the first few hundred inverse femtobarns of data.

The production of the charged $S^{\pm}$ is suppressed compared with the neutrals and the pseudoscalar $S_I^0$ does not have a significant branching ratio to the most promising detection channels $WW$ and $ZZ$. Hence we present in detail a discussion of the neutral scalar\footnote{Similar conclusions hold for the charged scalar and neutral pseudoscalar.}.  Expanding $H$ and $S$ about their vacuum expectation values we write
\begin{equation}
H^0 -\langle H^0 \rangle = {h^0 \over {\sqrt 2}},~~~~ S^0 -\langle S^0 \rangle = {S^0_R+iS_I^0 \over {\sqrt 2}}.
\end{equation}
The fields $h^0$ and $S_R^0$ mix and the resulting mass eigenstate fields $h$ and $S_R$ are approximately given by{\footnote{We assume that the parameters in the scalar potential are real so there is no $S_R^0-S_I^0$ mixing.}
\begin{equation}
\label{mixing}
 h \simeq h^0+{3 g_1v^2\over 2 M^2}S_R^0,~~~~~~S_R \simeq S_R^0-{3 g_1 v^2 \over2 M^2}h^0.
\end{equation}

The production rate for ${S_R}$ is very small. The dominant production mode is through
$b \, \bar{b} \to {S}_R$.  In this process the initial $b$ and $\bar{b}$ each come mostly from collinear gluon splitting and the remaining spectator $b$ quarks have very low transverse momentum to be observed in the final state. The large logarithms associated with collinear gluon splitting into light quark pairs leads to an enhancement of the rate by one or two orders of magnitude \cite{Dawson:2005vi,Dittmaier:2003ej,Dawson:2003kb} over $g\, g \to b \, \bar{b} \, S_R$ \footnote{This result is based on the NLO QCD calculation for $\bar{t} \, t + h$ production in  \cite{Dawson:2002tg,Reina:2001sf,Beenakker:2002nc,Beenakker:2001rj}.} where the
scalar is radiated off one of the final state $b$-quarks. This is because the final state $b$-quark which radiates the scalar is far offshell before emission and thus the rate does not receive the enhancement of large logarithms associated with collinear gluon splitting.
The cross-section for $b\, \bar{b}\to S_R$ at leading log takes the form
\begin{eqnarray}
\label{bbS}
\sigma(b \, \bar{b} \to {S}_R)_{\rm LHC} \simeq \frac{\eta_D^2   \, \pi} {3 \, s} \, \left(\frac{\tilde{m}_b^2}{v^2}\right) \, \int _{{M^2}/{s}}^1 \frac{dx}{x} b(x,\mu) \>\bar{b}(\frac{M^2}{x s},\mu),
\end{eqnarray}
where $b(x,\mu)$ and $\bar{b}(x,\mu)$ are the
$b$ quark and antiquark parton distribution functions respectively and $s$ is the center of mass energy squared. The large logs from collinear gluon splitting are summed into the parton distribution functions by choosing $\mu \sim M$. As seen from Eq.(\ref{bbS}), the $b \, \bar{b}\to {S}_R$ cross section receives an additional enhancement by a factor of $\eta_D^2  \sim 100$ compared to the production of a SM Higgs with the same mass. This production cross section as a function of the mass $M$ is shown in Fig.(\ref{production}) as the solid black curve. This curve was generated for the choice of $\eta_D=10$ and $g_1=0.5$. We see that at $M=1$ TeV the $b \, \bar{b} \to {S}_R$ cross section is about $10$ fb. Thus, for 100 fb$^{-1}$ of data one can expect the production of about 1000 neutral scalars $S_R$ from $b \, {\bar b}$ fusion. Note that this dominant production mechanism doesn't exist for the heavy charged scalars. 

\begin{figure} 
\centerline{\scalebox{1.0}{\includegraphics{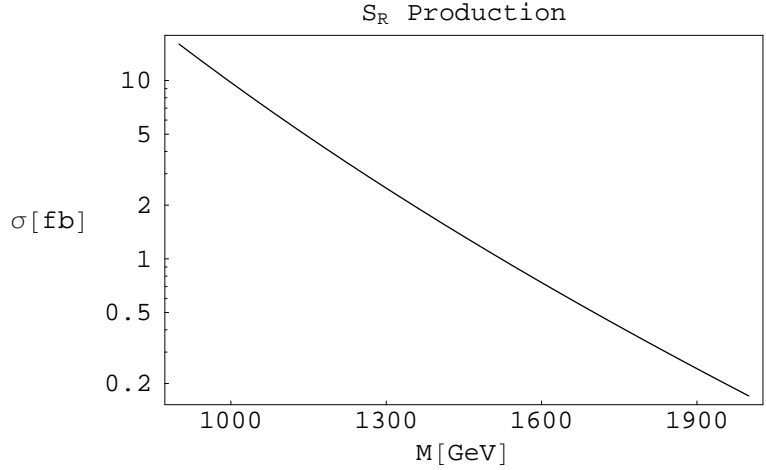}}}
\caption{The production cross section of $b \, \bar{b} \to {S}_R$ for the parameter choices of $\eta_D=10$ and $g_1=0.5$. The production cross section (for the same parameter choices) for $g \, g \to {S}_R$ is approximately 
$0.2 \, {\rm fb}$ for $M \sim 800 \, {\rm GeV}$ and falls quickly with increasing $M$ and thus is not shown.
The curve was generated using CTEQ5 parton distribution functions \cite{Lai:1999wy}. The $b$ and $t$ masses were evaluated at   $\mu= 1 \, {\rm TeV}$ using leading log running. We chose  $\Lambda_{QCD} = 0.1\,  {\rm GeV}$ and used the initial values ${m}_b = 4.26 \, {\rm GeV}$ determined from converting the result of the $1S$ fit to the 
b quark mass \cite{Bauer:2004ve} and the value $m_t = 170 \, {\rm GeV}$ from the PDG \cite{PDBook}}
\label{production}
\end{figure}

The next largest production mode of ${S}_R$ is through gluon fusion
and  is given by a direct modification of the SM cross section \cite{Dawson:1994ri}
\begin{eqnarray}
\label{ggS}
\sigma (g \, g\to {S}_R)_{\rm LHC} \simeq \frac{\alpha_s^2}{64\pi s } \frac{M^2}{v^2}\Bigg |\eta_D  \frac{\tilde{m}_b}{m_b} \,  {\rm I}\Big (\frac{M^2}{4 \, m_b^2}\Big ) -\frac{3g_1 v^2}{2M^2}  \,  {\rm I} \Big (\frac{M^2}{4 \, m_t^2}\Big ) \Bigg |^2\>  F[\mu,M,s],
\end{eqnarray}
where  we have used the functions
\begin{eqnarray}
F[\mu,M,s] = \int _{{M^2}/{s}}^1 \frac{dx}{x} g(x, \mu) g(\frac{M^2}{x s},\mu) ,
\end{eqnarray}
and  $ {\rm I}(y)$ for $y >1$ which is given by \cite{Bergstrom:1985hp}
\begin{equation}
{\rm I}(y)=  \frac{1}{2 \, y} + \frac{y-1}{2 \, y^2} \left[ i \, \pi \, \log \left(\sqrt{y} + \sqrt{y-1} \right) - \log^2 \left(\sqrt{y} + \sqrt{y-1} \right) + \frac{\pi^2}{4}  \right]. 
\label{if}
\end{equation}
Here $g(x,\mu)$ denotes the gluon parton distribution function.
As seen in Eq. (\ref{ggS}) this production channel receives significant contributions from  bottom and top loops. The bottom loop has a significant contribution due to the direct coupling of ${S}_R$ which involves $\eta_D \sim 10 \gg 1$. Although the direct coupling
of ${S}_R^0$ to the top quark is negligible for $\eta_U \ll 1$, the top loop still gives a significant contribution due to the mixing of ${S}_R^0$ with the Higgs $h^0$.
With the same parameters used as in Fig.(\ref{production}) at $M=1$ TeV one can expect the production via gluon fusion of only about 8 neutral scalars ${S}_R$ for 100 fb$^{-1}$ of data.

Other production mechanisms are similarly small. For example Higgs production (via vector-boson fusion) in association with massless jets, $q \, q \rightarrow q \, q \, h $, where $q=\{u,d,s\}$, is dominated by the Higgs being radiated off a virtual $W$ or $Z$ boson. So
\begin{equation}
{\sigma(q \, q \rightarrow q \, q \, {S_R}) \over \sigma(q \, q \rightarrow q \, q \, h)_{\rm SM}}\simeq \left({3g_1 v^2 \over 2 M^2}\right)^2 \ll 1.
\end{equation}

The pattern of possible decays of the new scalars in the model depends on the mass splittings between the various states. To simplify our discussion of the spectrum we neglect the $S_R^0-h^0$ mixing and assume that the coupling constants in the scalar potential are real. Then there is no $S_I^0-S_R^0$ mixing and the mass spectrum is,
\begin{eqnarray}
&&m^2_{S\pm} \simeq M^2+ g_2 \,  v^2, \nonumber \\
&&m^2_{S_R^0}\simeq M^2 + (g_2 + g_2' +g_2''/2) \, v^2, \nonumber \\
&&m^2_{S_I^0}\simeq M^2 + (g_2 - g_2' +g_2''/2) \, v^2.
\end{eqnarray}
We focus on the region of parameter space where the lightest scalar is $S_R^0$. For its decays it is important to include the effects of $S_R^0-h^0$ mixing. The most important decay modes of $S_R$ have the partial rates
\begin{eqnarray}
\Gamma( S_{R} \rightarrow {t \, \bar t})&\simeq&{27 g_1^2v^4\over 32 M^3\pi } \left({m_t \over v}\right)^2 , \\
\Gamma(S_{R} \rightarrow {b \, \bar b})&\simeq&{3|\eta_D|^2 M\over 8\pi } \left({\tilde{m_b} \over v}\right)^2 ,  \\ 
\Gamma( S_{R} \rightarrow {W^+ \,  W^-})&\simeq& \frac{g_1^2 v^2}{16 \pi M},  \\
\Gamma( S_{R} \rightarrow {Z \, Z})&\simeq&\frac{g_1^2 v^2}{32 \pi M},  \\
\Gamma (S_R \to h h) &\simeq & \frac{9 g_1^2 v^2}{32\pi M}, \\
\Gamma (S_R \to h h h) &\simeq & \frac{3 g_1^2 M}{1024 \pi ^3}.
\end{eqnarray}
\begin{figure} 
\centerline{\scalebox{1.0}{\includegraphics{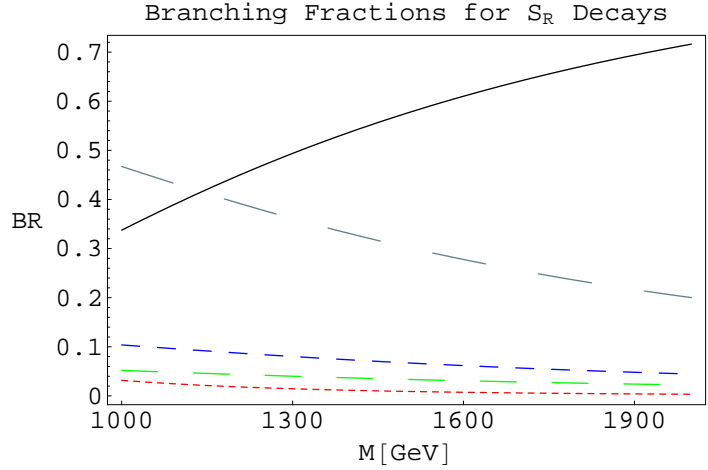}}}
\caption{Branching fractions for ${S}_R$ decays as a function of the it's mass $M$. The solid black curve denotes the branching ratio for $ S_{R} \rightarrow {b \, \bar b}$, the gray very-long-dashed curve denotes $S_R\to hh$, the red short-dashed curve is for $  S_{R} \rightarrow {t \, \bar t}$, the blue medium-dashed curve is for  $ S_{R} \rightarrow {W^+ W^-}$, and the green long-dashed curve is for ${S}_R\to Z^0 Z^0$.
We have not shown the curve for $S_R\to hhh$ in order to avoid too much clutter. These curves were generated with the parameter choices of $\eta_D=10$ and $g_1=0.5$.}
\label{bratios}
\end{figure}
 A plot of the branching fractions 
of ${S}_R$ as a function of the mass $M$ is shown in Fig.~(\ref{bratios}) for $g_1=0.5$, $M=1~{\rm TeV}$ and $\eta_D=10$. The dominant decay channels are $S_R\to hh$ and ${S}_R \to b \, \bar{b}$. The $S_R\to b\bar{b}$ channel is known to have  a large SM background. As we will discuss later, even the $S\to hh$ channel can be difficult to observe.   The final states where the ${S}_R$ decays to gauge bosons and at least one of the gauge bosons decays to electrons and/or muons have a cleaner experimental signature. Note that for the parameters used in Fig.~(\ref{bratios}) the total width of an  ${S}_R$ scalar of mass $1~{\rm TeV}$ is only 3~${\rm GeV}$. For comparison, note that the width of a standard model Higgs with a mass of $ 1 \, {\rm TeV}$ is about 700~${\rm GeV}$ \cite{Djouadi:2005gi}. This is because of the small vacuum expectation value of the heavy doublet which suppresses the coupling of ${S}_R $ to two gauge bosons.
\begin{table}
\begin{tabular}{|c||c|c|c|}
\hline
\text{Decay Channel} & $g_1=0.5, \eta_D=10$ &$g_1=1, \eta_D=5$ &$g_1=-2, \eta_D=20$ \\
\hline \hline
${S}_R\to Z^0Z^0\to \ell^+ \ell^-\ell^+\ell^-$ & $0.23$& $0.10$ & $8.0$ \\
\hline
${S}_R\to Z^0Z^0\to \ell^+ \ell^-\nu \bar{\nu}$ &1.4& 0.58 & 47\\
\hline
${S}_R\to W^+W^-\to \ell^+ \nu \ell^- \bar{\nu}$ &4.6& 2.0 & 160 \\
\hline
${S}_R\to W^+W^-\to  (\ell^+ \nu j \, j,  \ell^- \bar{\nu} j j)$ &3.0 $\times$ 10& 12 & 1.0 $\times $ 10$^3$ \\
\hline
\end{tabular}
\caption{Expected number of events for 100 \, fb$^{-1}$ of data at the LHC in the experimentally favored decay modes of ${S}_R$ for different choices of the parameters $g_1, \eta_D$. The production cross section is the sum of the $\sigma (g \, g\to {S}_R)_{\rm LHC}$ and $\sigma (b \, b\to {S}_R)_{\rm LHC}$ cross sections. We use
$\ell$ to denote either an electron or muon (i.e, we have summed over $l=e,\mu$) and $j$ denotes a single jet.   We have chosen the mass of ${S}_R$ at $M=1$ TeV and a center of mass energy of 14 TeV. After realistic selection cuts the final number of accepted events will be lower.}
\label{table1}
\end{table}

\begin{table}
\begin{tabular}{|c||c|c|c|}
\hline
\text{Decay Channel} & $g_1=0.5, \eta_D=10$ &$g_1=1, \eta_D=5$ &$g_1=-2, \eta_D=20$ \\
\hline \hline
${S}_R\to hh\to b\bar{b} \gamma \gamma$ & $1.1$& $0.46$ & $0.82$ \\
\hline
${S}_R\to hh\to b\bar{b} \tau^+\tau^-$ &34& 14& 26 \\
\hline
\end{tabular}
\caption{Expected number of events for 100\, fb$^{-1}$ of data at the LHC in the experimentally favored channels when $S_R$ first decays to a pair of light Higgses \cite{cms}. We have chosen the mass of $S_R$ at $M=1$ TeV and the Higgs mass at $m_h=120$ GeV.}
\label{table2}
\end{table}

In Table.~\ref{table1} we show the number of expected events 
for 100 fb$^{-1}$ of data at the LHC when $S_R$ decays to gauge bosons and $M=1$ TeV. For example with $g_1=0.5$, $M= 1~{\rm TeV}$ and $\eta_D=10$ we find that $\sigma(p p \rightarrow {S_R }X)Br( {S_R } \rightarrow W^+ \, W^-)Br(W^+ \, W^- \rightarrow \ell^+ \, \nu \, \ell^-  \bar{\nu_\ell}) \sim 0.05 ~{\rm fb}$, where we have summed over $l={e,\mu}$ leading to about five events with 100 fb$^{-1}$ of data.  For these parameters, the heavy scalar $S_R $ will not be detected at the LHC in the first few hundred femtobarns of integrated luminosity. In fact, within much of the region of parameter space where the coupling of the new S-doublet to charge $2/3$-quarks is suppressed (i.e., $\eta_U$  very small) the heavy scalar degrees of freedom associated with the doublet $S$ are difficult to detect at the LHC as shown in in the first two columns of Table.~\ref{table1}.  However, as seen in the third column of  Table.~\ref{table1}, even with $\eta_U$ small, there are regions of parameter space that are more promising for detection at LHC. For example, for $g_1=-2$, $M=1~{\rm TeV}$ and $\eta_D=20$ we find that $\sigma(p p \rightarrow {S_R }X)Br( { S_R } \rightarrow W^+ \,W^-)Br(W^+ \, W^- \rightarrow \ell^+ \, \nu \, \ell^-  \bar{\nu_\ell}) \sim 1.6~ {\rm fb}$ and detection of the new heavy scalar ${ S_R }$ at the LHC with a few hundred inverse femtobarns of integrated luminosity is more likely.

In Table.~\ref{table2} we show the number of expected events when $S_R$ decays to a pair of light Higgses and one of them decays to the experimentally favored $\gamma \gamma$ or $\tau^+ \tau^-$ channels. As seen in the table, for the parameters chosen detection is unlikely. The number of events 
in the last column are suppressed because for these parameters the Higgs decay rate to $b\bar{b}$ is enhanced by a factor of about 100 which reduces the Higgs branching ratio to $\gamma \gamma$ and $\tau^+ \tau^-$ by a similar factor.

\section{Concluding Remarks}
We have demonstrated that for regions of parameter space in multi doublet models the states of the 
new doublets are impossible to directly detect at LHC, using the first few hundred inverse femtobarns of data, and yet the effect of the new doublet on the total width of the 
light Higgs is very significant. In the simple two doublet model we considered in detail, the promising  $h \rightarrow \gamma \, \gamma$ and $h \rightarrow \tau^+ \tau^-$ signals at the LHC
for detecting a light Higgs could be significantly enhanced or suppressed. This demonstration emphasizes the 
importance of determining the properties of the Higgs boson in the presence of 
new physics that is difficult to directly detect at LHC. 

We thank Marat Gataullin for many helpful comments. This
work was supported in part by the DOE grant number DE-FG03-92ER40701
and DE-FG03-97ER40546.

\bibliographystyle{h-physrev3.bst}
\bibliography{Higgs}

\begin{thebibliography}{10}

\bibitem{Reina:2005ae}
L.~Reina,
\newblock (2005), hep-ph/0512377.

\bibitem{Glashow:1976nt}
S.~L. Glashow and S.~Weinberg,
\newblock Phys. Rev. {\bf D15}, 1958 (1977).

\bibitem{Cheng:1987rs}
T.~P. Cheng and M.~Sher,
\newblock Phys. Rev. {\bf D35}, 3484 (1987).

\bibitem{Luke:1993cy}
M.~E. Luke and M.~J. Savage,
\newblock Phys. Lett. {\bf B307}, 387 (1993), hep-ph/9303249.

\bibitem{Antaramian:1992ya}
A.~Antaramian, L.~J. Hall, and A.~Rasin,
\newblock Phys. Rev. Lett. {\bf 69}, 1871 (1992), hep-ph/9206205.

\bibitem{Atwood:1996vj}
D.~Atwood, L.~Reina, and A.~Soni,
\newblock Phys. Rev. {\bf D55}, 3156 (1997), hep-ph/9609279.

\bibitem{Chivukula:1987py}
R.~S. Chivukula and H.~Georgi,
\newblock Phys. Lett. {\bf B188}, 99 (1987).

\bibitem{Hall:1990ac}
L.~J. Hall and L.~Randall,
\newblock Phys. Rev. Lett. {\bf 65}, 2939 (1990).

\bibitem{D'Ambrosio:2002ex}
G.~D'Ambrosio, G.~F. Giudice, G.~Isidori, and A.~Strumia,
\newblock Nucl. Phys. {\bf B645}, 155 (2002), hep-ph/0207036.

\bibitem{Cirigliano:2005ck}
V.~Cirigliano, B.~Grinstein, G.~Isidori, and M.~B. Wise,
\newblock Nucl. Phys. {\bf B728}, 121 (2005), hep-ph/0507001.

\bibitem{Buras:2003jf}
A.~J. Buras,
\newblock Acta Phys. Polon. {\bf B34}, 5615 (2003), hep-ph/0310208.

\bibitem{Branco:2006hz}
G.~C. Branco, A.~J. Buras, S.~Jager, S.~Uhlig, and A.~Weiler,
\newblock (2006), hep-ph/0609067.

\bibitem{Manohar:2006gz}
A.~V. Manohar and M.~B. Wise,
\newblock Phys. Lett. {\bf B636}, 107 (2006), hep-ph/0601212.

\bibitem{Grinstein:2007iv}
B.~Grinstein and M.~Trott,
\newblock (2007), arXiv:0704.1505 [hep-ph].

\bibitem{Kile:2007ts}
J.~Kile and M.~J. Ramsey-Musolf,
\newblock (2007), arXiv:0705.0554 [hep-ph].

\bibitem{Graesser:2007yj}
M.~L. Graesser,
\newblock (2007), arXiv:0704.0438 [hep-ph].

\bibitem{Mantry:2007sj}
S.~Mantry, M.~J. Ramsey-Musolf, and M.~Trott,
\newblock (2007), arXiv:0707.3152 [hep-ph].

\bibitem{Pierce:2006dh}
A.~Pierce, J.~Thaler, and L.-T. Wang,
\newblock (2006), hep-ph/0609049.

\bibitem{Fox:2007in}
P.~J. Fox, Z.~Ligeti, M.~Papucci, G.~Perez, and M.~D. Schwartz,
\newblock (2007), arXiv:0704.1482 [hep-ph].

\bibitem{Rainwater:2007cp}
D.~Rainwater,
\newblock (2007), hep-ph/0702124.

\bibitem{Benedetti:2007sn}
D.~Benedetti {\em et~al.},
\newblock J. Phys. {\bf G34}, N221 (2007).

\bibitem{Cao:2007rm}
Q.-H. Cao, E.~Ma, and G.~Rajasekaran,
\newblock (2007), arXiv:0708.2939 [hep-ph].

\bibitem{Randall:2007as}
L.~Randall,
\newblock (2007), arXiv:0711.4360 [hep-ph].

\bibitem{Grinstein:1987pu}
B.~Grinstein and M.~B. Wise,
\newblock Phys. Lett. {\bf B201}, 274 (1988).

\bibitem{Grinstein:1987vj}
B.~Grinstein, R.~P. Springer, and M.~B. Wise,
\newblock Phys. Lett. {\bf B202}, 138 (1988).

\bibitem{Misiak:2006zs}
M.~Misiak {\em et~al.},
\newblock Phys. Rev. Lett. {\bf 98}, 022002 (2007), hep-ph/0609232.

\bibitem{Misiak:2006ab}
M.~Misiak and M.~Steinhauser,
\newblock Nucl. Phys. {\bf B764}, 62 (2007), hep-ph/0609241.

\bibitem{Aubert:2005cua}
BABAR, B.~Aubert {\em et~al.},
\newblock Phys. Rev. {\bf D72}, 052004 (2005), hep-ex/0508004.

\bibitem{Aubert:2006gg}
BaBar, B.~Aubert {\em et~al.},
\newblock Phys. Rev. Lett. {\bf 97}, 171803 (2006), hep-ex/0607071.

\bibitem{Koppenburg:2004fz}
Belle, P.~Koppenburg {\em et~al.},
\newblock Phys. Rev. Lett. {\bf 93}, 061803 (2004), hep-ex/0403004.

\bibitem{Barberio:2006bi}
Heavy Flavor Averaging Group (HFAG), E.~Barberio {\em et~al.},
\newblock (2006), hep-ex/0603003.

\bibitem{Becirevic:2006nm}
D.~Becirevic, V.~Lubicz, and F.~Mescia,
\newblock Nucl. Phys. {\bf B769}, 31 (2007), hep-ph/0611295.

\bibitem{Ball:2006eu}
P.~Ball, G.~W. Jones, and R.~Zwicky,
\newblock Phys. Rev. {\bf D75}, 054004 (2007), hep-ph/0612081.

\bibitem{Acosta:2005bk}
CDF, D.~Acosta {\em et~al.},
\newblock Phys. Rev. {\bf D72}, 072004 (2005), hep-ex/0506042.

\bibitem{Belotsky:2002ym}
K.~Belotsky, D.~Fargion, M.~Khlopov, R.~Konoplich, and K.~Shibaev,
\newblock Phys. Rev. {\bf D68}, 054027 (2003), hep-ph/0210153.

\bibitem{BahatTreidel:2006kx}
O.~Bahat-Treidel, Y.~Grossman, and Y.~Rozen,
\newblock JHEP {\bf 05}, 022 (2007), hep-ph/0611162.

\bibitem{Dawson:2005vi}
S.~Dawson, C.~B. Jackson, L.~Reina, and D.~Wackeroth,
\newblock Mod. Phys. Lett. {\bf A21}, 89 (2006), hep-ph/0508293.

\bibitem{Dittmaier:2003ej}
S.~Dittmaier, M.~Kramer, and M.~Spira,
\newblock Phys. Rev. {\bf D70}, 074010 (2004), hep-ph/0309204.

\bibitem{Dawson:2003kb}
S.~Dawson, C.~B. Jackson, L.~Reina, and D.~Wackeroth,
\newblock Phys. Rev. {\bf D69}, 074027 (2004), hep-ph/0311067.

\bibitem{Dawson:2002tg}
S.~Dawson, L.~H. Orr, L.~Reina, and D.~Wackeroth,
\newblock Phys. Rev. {\bf D67}, 071503 (2003), hep-ph/0211438.

\bibitem{Reina:2001sf}
L.~Reina and S.~Dawson,
\newblock Phys. Rev. Lett. {\bf 87}, 201804 (2001), hep-ph/0107101.

\bibitem{Beenakker:2002nc}
W.~Beenakker {\em et~al.},
\newblock Nucl. Phys. {\bf B653}, 151 (2003), hep-ph/0211352.

\bibitem{Beenakker:2001rj}
W.~Beenakker {\em et~al.},
\newblock Phys. Rev. Lett. {\bf 87}, 201805 (2001), hep-ph/0107081.

\bibitem{Lai:1999wy}
CTEQ, H.~L. Lai {\em et~al.},
\newblock Eur. Phys. J. {\bf C12}, 375 (2000), hep-ph/9903282.

\bibitem{Bauer:2004ve}
C.~W. Bauer, Z.~Ligeti, M.~Luke, A.~V. Manohar, and M.~Trott,
\newblock Phys. Rev. {\bf D70}, 094017 (2004), hep-ph/0408002.

\bibitem{PDBook}
W.-M. {Yao} {\em et~al.},
\newblock {Journal of Physics G} {\bf 33}, 1+ (2006).

\bibitem{Dawson:1994ri}
S.~Dawson,
\newblock (1994), hep-ph/9411325.

\bibitem{Bergstrom:1985hp}
L.~Bergstrom and G.~Hulth,
\newblock Nucl. Phys. {\bf B259}, 137 (1985).

\bibitem{Djouadi:2005gi}
A.~Djouadi,
\newblock (2005), hep-ph/0503172.

\bibitem{cms}
CMS Collaboration, G.~Bayatian {\em et~al.},
\newblock  {\bf CERN/LHCC 2006-021} (2006).

\end{thebibliography}

\end{document}